\documentstyle[aps]{revtex}

\newcommand{\bel}[1]{\begin{equation}\label{#1}}
\newcommand{\be}{\begin{equation}}
\newcommand{\ee}{\end{equation}}

\newcommand{\beal}[1]{\begin{eqnarray}\label{#1}}
\newcommand{\bea}{\begin{eqnarray}}
\newcommand{\eea}{\end{eqnarray}}

\newcommand{\bean}{\begin{eqnarray*}}
\newcommand{\eean}{\end{eqnarray*}}

\newcommand{\ba}{\begin{array}}
\newcommand{\ea}{\end{array}}

\newcommand{\bab}{\begin{abstract}}
\newcommand{\eab}{\end{abstract}}

\newcommand{\bml}{\begin{mathletters}}
\newcommand{\eml}{\end{mathletters}}

\newcommand{\q}{\quad}
\newcommand{\qq}{\quad\quad}

\newcommand{\bfm}[1]{\mbox{\boldmath $#1$}}

\newcommand{\dv}{\partial}

\newcommand{\bam}{\left( \begin{array}}
\newcommand{\eam}{\end{array} \right)}

\newcommand{\baq}[4]{\left( \begin{array}{c}{#1}\\{#2}\\{#3}\\
{#4}\end{array} \right)}

\newcommand{\bamd}[4]{\left( \begin{array}{cc}{#1}&{#2}\\
{#3}&{#4}\end{array} \right)}

\newcommand{\bamq}[4]{\left( \begin{array}{cccc}{#1}&{#2}&{#3}&{#4}\\}
\newcommand{\bamc}[5]{\left( \begin{array}{ccccc}{#1}&{#2}&{#3}&{#4}&{#5}\\}

\newcommand{\raw}{\rightarrow}

\newcommand{\lrw}{\leftrightarrow}

\newcommand{\cg}{\gamma}
\newcommand{\dg}{\delta}
\newcommand{\eg}{\epsilon}

\newcommand{\sg}{\sigma}

\newcommand{\vpg}{\varphi}

\def\xc{{\mathchoice {\setbox0=\hbox{$\displaystyle\rm C$}\hbox{\hbox
to0pt{\kern0.4\wd0\vrule height0.9\ht0\hss}\box0}}
{\setbox0=\hbox{$\textstyle\rm C$}\hbox{\hbox
to0pt{\kern0.4\wd0\vrule height0.9\ht0\hss}\box0}}
{\setbox0=\hbox{$\scriptstyle\rm C$}\hbox{\hbox
to0pt{\kern0.4\wd0\vrule height0.9\ht0\hss}\box0}}
{\setbox0=\hbox{$\scriptscriptstyle\rm C$}\hbox{\hbox
to0pt{\kern0.4\wd0\vrule height0.9\ht0\hss}\box0}}}}

\def\xg{{\mathchoice {\setbox0=\hbox{$\displaystyle\rm G$}\hbox{\hbox
to0pt{\kern0.4\wd0\vrule height0.9\ht0\hss}\box0}}
{\setbox0=\hbox{$\textstyle\rm G$}\hbox{\hbox
to0pt{\kern0.4\wd0\vrule height0.9\ht0\hss}\box0}}
{\setbox0=\hbox{$\scriptstyle\rm G$}\hbox{\hbox
to0pt{\kern0.4\wd0\vrule height0.9\ht0\hss}\box0}}
{\setbox0=\hbox{$\scriptscriptstyle\rm G$}\hbox{\hbox
to0pt{\kern0.4\wd0\vrule height0.9\ht0\hss}\box0}}}}

\def\xi{{\rm I\!I}}

\def\xo{{\mathchoice {\setbox0=\hbox{$\displaystyle\rm O$}\hbox{\hbox
to0pt{\kern0.4\wd0\vrule height0.9\ht0\hss}\box0}}
{\setbox0=\hbox{$\textstyle\rm O$}\hbox{\hbox
to0pt{\kern0.4\wd0\vrule height0.9\ht0\hss}\box0}}
{\setbox0=\hbox{$\scriptstyle\rm O$}\hbox{\hbox
to0pt{\kern0.4\wd0\vrule height0.9\ht0\hss}\box0}}
{\setbox0=\hbox{$\scriptscriptstyle\rm O$}\hbox{\hbox
to0pt{\kern0.4\wd0\vrule height0.9\ht0\hss}\box0}}}}

\def\xq{{\mathchoice {\setbox0=\hbox{$\displaystyle\rm
Q$}\hbox{\raise
0.15\ht0\hbox to0pt{\kern0.4\wd0\vrule height0.8\ht0\hss}\box0}}
{\setbox0=\hbox{$\textstyle\rm Q$}\hbox{\raise
0.15\ht0\hbox to0pt{\kern0.4\wd0\vrule height0.8\ht0\hss}\box0}}
{\setbox0=\hbox{$\scriptstyle\rm Q$}\hbox{\raise
0.15\ht0\hbox to0pt{\kern0.4\wd0\vrule height0.7\ht0\hss}\box0}}
{\setbox0=\hbox{$\scriptscriptstyle\rm Q$}\hbox{\raise
0.15\ht0\hbox to0pt{\kern0.4\wd0\vrule height0.7\ht0\hss}\box0}}}}

\def\xs{{\mathchoice
{\setbox0=\hbox{$\displaystyle     \rm S$}\hbox{\raise0.5\ht0\hbox
to0pt{\kern0.35\wd0\vrule height0.45\ht0\hss}\hbox
to0pt{\kern0.55\wd0\vrule height0.5\ht0\hss}\box0}}
{\setbox0=\hbox{$\textstyle        \rm S$}\hbox{\raise0.5\ht0\hbox
to0pt{\kern0.35\wd0\vrule height0.45\ht0\hss}\hbox
to0pt{\kern0.55\wd0\vrule height0.5\ht0\hss}\box0}}
{\setbox0=\hbox{$\scriptstyle      \rm S$}\hbox{\raise0.5\ht0\hbox
to0pt{\kern0.35\wd0\vrule height0.45\ht0\hss}\raise0.05\ht0\hbox
to0pt{\kern0.5\wd0\vrule height0.45\ht0\hss}\box0}}
{\setbox0=\hbox{$\scriptscriptstyle\rm S$}\hbox{\raise0.5\ht0\hbox
to0pt{\kern0.4\wd0\vrule height0.45\ht0\hss}\raise0.05\ht0\hbox
to0pt{\kern0.55\wd0\vrule height0.45\ht0\hss}\box0}}}}

\def\xt{{\mathchoice {\setbox0=\hbox{$\displaystyle\rm
T$}\hbox{\hbox to0pt{\kern0.3\wd0\vrule height0.9\ht0\hss}\box0}}
{\setbox0=\hbox{$\textstyle\rm T$}\hbox{\hbox
to0pt{\kern0.3\wd0\vrule height0.9\ht0\hss}\box0}}
{\setbox0=\hbox{$\scriptstyle\rm T$}\hbox{\hbox
to0pt{\kern0.3\wd0\vrule height0.9\ht0\hss}\box0}}
{\setbox0=\hbox{$\scriptscriptstyle\rm T$}\hbox{\hbox
to0pt{\kern0.3\wd0\vrule height0.9\ht0\hss}\box0}}}}

\def\xz{{\mathchoice {\hbox{$\sf\textstyle Z\kern-0.4em Z$}}
{\hbox{$\sf\textstyle Z\kern-0.4em Z$}}
{\hbox{$\sf\scriptstyle Z\kern-0.3em Z$}}
{\hbox{$\sf\scriptscriptstyle Z\kern-0.2em Z$}}}}

\newcommand{\fs}{\footnotesize}

\newcommand{\bii}{\begin{itemize}}
\newcommand{\eii}{\end{itemize}}

\newcommand{\ben}{\begin{enumerate}}
\newcommand{\een}{\end{enumerate}}

\newcommand{\bq}{\begin{quote}}
\newcommand{\eq}{\end{quote}}

\newcommand{\bc}{\begin{center}}
\newcommand{\ec}{\end{center}}

\newcommand{\btb}{\begin{table}}
\newcommand{\etb}{\end{table}}

\newcommand{\bt}{\begin{tabular}}
\newcommand{\et}{\end{tabular}}

\newcommand{\br}{\begin{flushright}}
\newcommand{\er}{\end{flushright}}

\newcommand{\bl}{\begin{flushleft}}
\newcommand{\el}{\end{flushleft}}

\newcommand{\bref}{\begin{references}}
\newcommand{\eref}{\end{references}}

\newcommand{\bb}{\begin{thebibliography}{99}}
\newcommand{\eb}{\end{thebibliography}}
\newcommand{\bi}{\bibitem}

\newcommand{\btp}{\begin{titlepage}}
\newcommand{\etp}{\end{titlepage}}

\newcommand{\go}{\section{Introduction}}
\newcommand{\con}{\section{Conclusions}}
\newcommand{\ack}{\section*{Acknowledgments}}

\newcommand{\ixa}[3]{Int.~J.~Mod.~Phys.~A               {\bf #1},  #2  (19#3)}

\newcommand{\jxe}[3]{J.~Math.~Phys.                     {\bf #1},  #2  (19#3)}

\newcommand{\mxb}[3]{Mod.~Phys.~Lett.~A                 {\bf #1},  #2  (19#3)}

\newcommand{\nxd}[3]{Nuovo Cimento                      {\bf #1},  #2  (19#3)}

\newcommand{\pxf}[3]{Phys.~Rev.~D                       {\bf #1},  #2  (19#3)}

\newcommand{\pxxa}[3]{Prog.~Theor.~Phys.                {\bf #1},  #2  (19#3)}

\newcommand{\xxx}[3]{{\bf #1},  #2  (19#3)}

\newcommand{\co}{\mbox{\boldmath $\cal C$}}
\newcommand{\qu}{\mbox{\boldmath $\cal H$}}
\newcommand{\oct}{\mbox{\boldmath $\cal O$}}
\newcommand{\rea}{\mbox{\boldmath $\cal R$}}
\newcommand{\glr}{GL(8, \rea )}
\newcommand{\glc}{GL(4,  \co )}

\title{OCTONIONIC REPRESENTATIONS OF GL(8,${\cal R}$) AND 
GL(4, ${\cal C}$)}

\author{Stefano De Leo\thanks{{\sl deleos@le.infn.it}}$^{(a,b)}$ and 
Khaled Abdel-Khalek\thanks{{\sl khaled@le.infn.it}}$^{(a)}$} 
\address{$^{(a)}$~Dipartimento di Fisica, Universit\`a di Lecce\\
$^{(b)}$~Istituto Nazionale di Fisica Nucleare, Sezione di Lecce\\
- Lecce, 73100, Italy -}

\date{Revised version - June, 1996}

\draft

\begin{document}

\maketitle

\bab
Octonionic algebra being nonassociative is difficult to manipulate. We 
introduce left-right octonionic barred operators which enable us to 
reproduce the associative $GL(8,{\cal R})$ group. Extracting the basis of 
$GL(4, {\cal C})$,  we establish an interesting connection between the 
structure of left-right octonionic barred  operators and generic $4\times 4$ 
complex matrices. As an application we give an octonionic representation of 
the 4-dimensional Clifford algebra.
\eab

\renewcommand{\thefootnote}{\sharp\arabic{footnote}}

\go

Semi-simple Lie groups, classified in four categories: orthogonal 
groups, unitary groups, symplectic groups and exceptional groups, were 
respectively associated with real, complex, quaternionic and octonionic 
algebras. Thus, such algebras became the core of the classification of 
possible symmetries in physics~\cite{om1,om2,om3,om4}.

We know that the antihermitian generators of $SU(2, \co)$ can be 
represented by the three quaternionic imaginary units $e_1$, $e_2$, $e_3$
\be
e_1 \lrw \left( \begin{array}{cc} i & 0\\ 0 & $-$i \end{array} \right)
\q , \q 
e_2 \lrw \left( \begin{array}{cc} 0 & $-$1\\ 1 & 0 \end{array} \right)
\q , \q 
e_3 \lrw \left( \begin{array}{cc} 0 & $-$i\\ $-$i  & 0 \end{array} \right)
\q .
\ee
It permits any quaternionic numbers or matrix to be translated into a 
complex matrix but {\tt not} necessarily vice-versa. In fact, to 
define the most general $2\times 2$ complex matrix, we need 8 real numbers. 
This problem is solved by introducing the barred quaternion 
$1\mid e_1$ ($\lrw i \openone_{2\times 2}$) which allows to obtain a 
faithful quaternionic representation of $GL(2, \co)$~\cite{qua2}. 

Exploiting the barred operator idea, we find the following 16  
quaternionic operators
\be
1 \q , \q {\bf Q} \q , \q 1 \mid {\bf Q} \q , \q e_1 \mid {\bf Q} \q , \q 
e_2 \mid {\bf Q} \q , \q e_3 \mid {\bf Q} \q ,
\ee
where ${\bf Q}\equiv (e_1, \; e_2, \; e_3)$. These operators become 
essential  
to formulate special relativity with real quaternions~\cite{rel}, allowing 
to overcome the difficulties which in the past did not permit a (real) 
quaternionic version of special relativity. Besides, they can be used 
to give a representation of $GL(4, \rea)$. The situation can be summarized 
as follows
\bean
GL(2, \co)~ & ~~\lrw ~~& q + p \mid e_1 \q ,\\
GL(4, \rea) & ~~\lrw ~~& q + p \mid e_1+ r \mid e_2+ s \mid e_3 \q ,
\eean
with $q, \; p, \; r, \; s$ quaternionic numbers.

Inspiring by this sequence we try to extend it and find an isomorphism 
between octonions and  $8\times 8$ real [or $4\times 
4$ complex] matrices. 
Obviously a first difficulty is the following: 
The octonionic algebra is nonassociative whereas $GL(8,\rea)$ 
[or $GL(4,\co)$], satisfying the Jacobi identity, is 
associative. This seems a hopeless situation.

In this paper, we introduce left/right octonionic barred operators which 
enables us to find translation rules between $8\times 8$ real matrices and 
octonionic numbers. On our road we also find an interesting isomorphism between 
the structure of left/right octonionic barred operators, on the one hand, 
and $4\times 4$ complex matrices, on the other hand. 

This article is organized as follows: In section II, we give a brief 
introduction to the octonionic division algebra. In section III, we  
discuss octonionic barred operators and explain the need to distinguish 
between left-barred and right-barred operators. In section IV, 
we investigate the 
relation between barred octonions and $8\times 8$ real matrices. 
In this section, we also give the translation rules 
between our octonionic barred operators and $\glc$ and as an application we 
write down  octonionic representations of the 4-dimensional Clifford algebra. 
Two appendices, containing explicit octonionic representation of $\glr$ and 
$\glc$, are included. Our conclusions and future developments are drawn in 
the final section.

\section{Octonionic algebra}

A remarkable theorem of Albert~\cite{alb2} shows that the only algebras, 
$\cal A$, over the reals, with unit element and admitting a real modulus 
function 
$N(a)$ ($a \in {\cal A}$) with the following properties
\bml
\bea
N(0) & = & 0 \q ,\\
N(a) & >& 0 \qq \mbox{if} \q a\neq 0 \q ,\\
N(ra) & = & \vert r \vert ~N(a) \qq (~r \in \rea~) \q ,\\
N(a_{1}a_{2}) & \leq & N(a_{1})+N(a_{2}) \q ,
\eea
\eml
are the reals, $\rea$, the complex, $\co$, the quaternions,   
$\qu$ ($\qu$ in honour of Hamilton~\cite{ham}), and the octonions, $\oct$   
(or Graves-Cayley numbers~\cite{gra,cay}). Albert's theorem generalizes 
famous nineteenth-century results of Frobenius~\cite{fro} and 
Hurwitz~\cite{hur}, who first reached the same conclusion but with the 
additional assumption that $N(a)^{2}$ is a quadratic form.

In addition to Albert's theorem on algebras admitting a modulus function 
$N(a)$, we can characterize the algebras $\rea$, $\co$, $\qu$ and 
$\oct$ by the concept of {\tt division algebra} (in which one has no nonzero 
divisors of zero). A classical theorem~\cite{bot,ker} states that the only 
division algebras over the reals are algebras of dimensions 1, 2, 4 and 
8, the only associative division algebras over the reals are $\rea$, $\co$ 
and $\qu$, whereas the  {\tt nonassociative} algebras include the octonions 
$\oct$ (an interesting discussion concerning nonassociative algebras is 
presented 
in~\cite{oku}). For a very nice review of aspects of the quaternionic and 
octonionic algebras see ref.~\cite{gur1} and the recent book of 
Adler~\cite{adl}.
In this paper we will deal with octonions and their generalizations.

We now summarize our notation for the octonionic algebra and 
introduce useful elementary properties to manipulate the nonassociative 
numbers. There is a number of equivalent ways to represent the octonions 
multiplication table. Fortunately, it is always possible to choose an 
orthonormal basis $(e_0 , \ldots ,e_7 )$ such that
\be
{\cal O} = r_{0}+\sum_{m=1}^{7} r_{m}e_{m} \qq (~r_{0,...,7}~~ \mbox{reals}~) \q , 
\ee 
where $e_{m}$ are elements obeying the noncommutative and nonassociative 
algebra
\be
e_{m}e_{n}=-\dg_{mn}+ \eg_{mnp}e_{p} \qq 
(~\mbox{{\fs $m, \; n, \; p =1,..., 7$}}~) 
\q ,
\ee
with $\eg_{mnp}$ totally antisymmetric and equal to unity for the seven 
combinations 
\[
123, \; 145, \; 176, \; 246, \; 257, \; 347 \; \mbox{and} \; 365  
\]
(each cycle represents a 
quaternionic subalgebra). The norm, $N({\cal O})$, for the octonions is defined by
\be
N({\cal O})=({\cal O}^{\dag}{\cal O})^{\frac{1}{2}}=
({\cal O}{\cal O}^{\dag})^{\frac{1}{2}}=
(r_{0}^{2}+  ... + r_{7}^{2})^{\frac{1}{2}} \q ,
\ee
with the octonionic conjugate ${\cal O}^{\dag}$ given by
\be
{\cal O}^{\dag}=r_{0}-\sum_{m=1}^{7} r_{m}e_{m} \q . 
\ee 
The inverse is then 
\be
{\cal O}^{-1}={\cal O}^{\dag}/N({\cal O}) \qq (~{\cal O}\neq 0~) \q .
\ee
We can define an {\tt associator} (analogous to the usual algebraic 
commutator) as follows
\bel{ass}
\{x, \; y , \; z\}\equiv (xy)z-x(yz) \q ,
\ee
where, in each term on the right-hand, we must, first of all, perform 
the multiplication in brackets. 
Note that for real, complex and quaternionic numbers the associator is 
trivially null. For octonionic imaginary units we have
\bel{eqass}
\{e_{m}, \; e_{n}, \; e_{p} \}\equiv(e_{m}e_{n})e_{p}-e_{m}(e_{n}e_{p})=
2 \eg_{mnps} e_{s} \q ,
\ee
with $\eg_{mnps}$ totally antisymmetric and equal to unity for the seven 
combinations 
\[ 
1247, \; 1265, \; 2345, \; 2376, \; 3146, \; 3157 \; \mbox{and} \; 4567 \q .
\]
Working with octonionic numbers the associator~(\ref{ass}) is in general 
non-vanishing, however, the ``alternative condition'' is fulfilled
\bel{rul}
\{ x, \; y, \; z\}+\{ z, \; y, \; x\}=0 \q .
\ee

\section{Left/Right-barred operators}

In 1989, writing a quaternionic Dirac equation~\cite{dir2}, Rotelli introduced
a {\tt barred}  momentum operator
\be
-\bfm{\dv}\mid i \qq [~(-\bfm{\dv}\mid i)\psi\equiv -\bfm{\dv}\psi i~] \q .
\ee
In recent papers~\cite{deleos}, {\tt partially barred quaternions} 
\be
q+p\mid i \qq [~q, \; p \in \qu~] \q ,
\ee
have been used to formulate a quaternionic quantum mechanics and field 
theory. From the viewpoint of group structure, these barred numbers are very 
similar to complexified quaternions~\cite{mor2}
\be
q+{\cal I}p
\ee
(the imaginary unit ${\cal I}$ commutes with the quaternionic 
imaginary units $i, \; j, \; k$), but in physical problems, like eigenvalue 
calculations, tensor 
products, relativistic equations solutions, they give different results. 

A complete generalization for quaternionic numbers is represented by 
the following barred operators
\be
q_{1} + q_{2}\mid i + q_{3}\mid j + q_{4}\mid k \qq
[~q_{1,...,4} \in \qu~] \q ,
\ee
which we call {\tt fully barred quaternions}, or simply barred 
quaternions. They, with their 16 linearly 
independent elements, form
a basis of $GL(4, \rea )$. They are successfully used 
to reformulate Lorentz space-time transformations~\cite{rel} and write down a 
one-component Dirac equation~\cite{dir4}.

Thus, it seems to us natural to investigate the existence of
{\tt barred octonions}
\be
{\cal O}_{0}+ \sum_{m=1}^{7} {\cal O}_{m}\mid e_{m} \qq
[~{\cal O}_{0, ...,7}~~ \mbox{octonions}~] \q  .
\ee
Nevertheless, we must observe that an octonionic {\tt barred} operator, 
\bfm{a\mid b}, which acts on octonionic wave functions, $\psi$, 
\[ [~a\mid b~]~\psi \equiv a\psi b \q , \]
is not a well defined object. For $a\neq b$ the triple product $a\psi b$ 
could be either $(a\psi)b$ or $a(\psi b)$. So, in order to avoid the 
ambiguity due to the nonassociativity of 
the octonionic numbers, we need to define left/right-barred 
operators. We will indicate {\tt left-barred} operators by 
\bfm{a~)~b}, with $a$ and $b$ which represent octonionic numbers. They 
act on octonionic functions $\psi$ as follows
\bml
\be
[~a~)~b~]~\psi = (a\psi)b \q .
\ee
In similar way we can introduce {\tt right-barred} operators, defined by 
\bfm{a~(~b} ,
\be
[~a~(~b~]~\psi = a(\psi b) \q .
\ee
\eml
Obviously, there are barred-operators in which the nonassociativity is not 
of relevance, like 
\[ 1~)~a = 1~(~a \equiv 1\mid a \q . \]
Furthermore, from eq.~(\ref{rul}), we have
\[ \{ x, \; y, \; x\}=0 \q ,\]
so  
\[ a~)~a = a~(~a  \equiv a\mid a \q .\]
At first glance it seems that we must consider the following 
106 barred-operators:
\bc
\bt{lr}
$1, ~ e_{m}, ~ 1\mid e_{m}$ & {\fs ~~~~~(15 elements)} ,\\
$e_{m}\mid e_{m}$ &{\fs (7)} ,\\
$e_{m}~)~e_{n}$~~~~{\fs $(m\neq n)$} & {\fs (42)} ,\\
$e_{m}~(~e_{n}$~~~~{\fs $(m\neq n)$} & {\fs (42)} ,\\
{\fs $(m, \; n =1, ..., 7) \q .$} & 
\et
\ec
Nevertheless, it is possible to prove that each right-barred operator can be 
expressed by a suitable combination of left-barred operators. For example, 
from eq.~(\ref{rul}), by posing $x=e_{m}$ and $z=e_{n}$, we quickly 
obtain 
\bel{f1}
e_{m}~(~e_{n} + e_{n}~(~e_{m} ~~\equiv ~~e_{m}~)~e_{n} + e_{n}~)~e_{m} \q .
\ee
So we can represent the most general octonionic operator by only left-barred 
objects 
\bel{go}
{\cal O}_{0}+\sum_{m=1}^{7} {\cal O}_{m}~)~e_{m} \qq 
[~{\cal O}_{0, ...,7}~~ \mbox{octonions}~] \q ,
\ee
reducing to 64 the previous 106 elements. This suggests a 
correspondence between our barred  octonions~(\ref{go}) and 
$GL(8, \rea)$ (a complete discussion about the 
above-mentioned relationship is given in the following section).

\section{Translation Rules}

The nonassociativity of octonions represents a challenge.
We overcome the problems due to the octonions nonassociativity by introducing 
left/right-barred operators. We 
discuss in the next subsection their relation to $GL(8, \rea)$. In that  
subsection, we present our translation idea and give some explicit examples 
which allow us to establish the isomorphism between our octonionic left/right 
barred operators and $GL(8, \rea)$. In subsection IV-b, we focus our 
attention on the group $GL(4, \co) \subset GL(8, \rea)$. In doing so, 
we find that only particular combinations of octonionic barred operators 
give us suitable candidates for the $GL(4, \co)$-translation. Finally, 
in subsection IV-c, we explicitly give  two octonionic 
representations for 
the Dirac gamma-matrices (and consequently we are able to write down, for 
the first time,  octonionic representations for the 4-dimensional Clifford 
algebra).

\subsection*{IV-a. Relation between barred operators and $8\times 8$ real 
matrices}

In order to explain the idea of translation, let us look 
explicitly at the action of the operators $1\mid e_1$ and $e_2$,  on a generic 
octonionic  function $\vpg$
\be
\vpg = \vpg_0
 + e_1 \vpg_1 + e_2 \vpg_2 + e_3 \vpg_3
+ e_4 \vpg_4 + e_5 \vpg_5 + e_6 \vpg_6 + e_7 \vpg_7
\q [~\vpg_{0,\dots ,7} \in \rea~] \q .
\ee
We have
\bml
\beal{opa}
[~1\mid e_{1}~]~\vpg ~ \equiv ~\vpg e_1 & ~=~ & 
e_1 \vpg_0 - \vpg_1 - e_3 \vpg_2 + e_2 \vpg_3
- e_5 \vpg_4 + e_4 \vpg_5 + e_7 \vpg_6 - e_6 \vpg_7 \q , \\
 e_{2}\vpg & ~=~ & e_2 \vpg_0 - e_3 \vpg_1 - \vpg_2 + e_1 \vpg_3
+ e_6 \vpg_4 + e_7 \vpg_5 - e_4 \vpg_6 - e_5 \vpg_7
\q .
\eea
\eml
If we represent our octonionic function $\vpg$ by the following real column 
vector
\be
\vpg ~ \lrw ~ \left( \begin{array}{c}
\vpg_0\\
\vpg_1\\
\vpg_2\\
\vpg_3\\
\vpg_4\\
\vpg_5\\
\vpg_6\\
\vpg_7
\end{array}
\right) \q ,
\ee
we can rewrite  eqs.~(\ref{opa}-b) in matrix form,
\bml
\bea
\left(
\begin{array}{cccccccc}
0 & $-1$ & 0 & 0 & 0 & 0 & 0 &0\\
1 & 0 & 0 & 0 & 0 & 0 & 0 &0\\
0 & 0 & 0 & 1 & 0 & 0 & 0 &0\\
0 & 0 & $-1$& 0 & 0 & 0 & 0 &0\\
0 & 0 & 0 & 0 & 0 & 1 & 0 &0\\
0 & 0 & 0 & 0 &$ -1$& 0 & 0 &0\\
0 & 0 & 0 & 0 & 0 & 0 & 0 &$-1$\\
0 & 0 & 0 & 0 & 0 & 0 & 1 &0
\end{array}
\right)
\left( \begin{array}{c}
\vpg_0\\
\vpg_1\\
\vpg_2\\
\vpg_3\\
\vpg_4\\
\vpg_5\\
\vpg_6\\
\vpg_7
\end{array}
\right) & = &
\left( \begin{array}{c}
$-$\vpg_1\\
\vpg_0\\
\vpg_3\\
$-$\vpg_2\\
\vpg_5\\
$-$\vpg_4\\
$-$\vpg_7\\
\vpg_6
\end{array} \right) \q , \\
\left(
\begin{array}{cccccccc}
0 & 0 &$-1$ & 0 & 0 & 0 & 0 &0\\
0 & 0 & 0 & 1 & 0 & 0 & 0 &0\\
1 & 0 & 0 & 0 & 0 & 0 & 0 &0\\
0 & $-1$& 0 & 0 & 0 & 0 & 0 &0\\
0 & 0 & 0 & 0 & 0 & 0 & $-1$&0\\
0 & 0 & 0 & 0 & 0 & 0 & 0 &$-1$\\
0 & 0 & 0 & 0 & 1 & 0 & 0 &0\\
0 & 0 & 0 & 0 & 0 & 1 & 0 &0
\end{array}
\right)
\left( \begin{array}{c}
\vpg_0\\
\vpg_1\\
\vpg_2\\
\vpg_3\\
\vpg_4\\
\vpg_5\\
\vpg_6\\
\vpg_7
\end{array}
\right) & = &
\left( \begin{array}{c}
$-$\vpg_2\\
\vpg_3\\
\vpg_0\\
$-$\vpg_1\\
$-$\vpg_6 \\
$-$\vpg_7\\
\vpg_4\\
\vpg_5
\end{array} \right) \q . 
\eea
\eml
In this way we can immediately obtain a real matrix representation for the 
octonionic barred operators $1\mid e_{1}$ and $e_{2}$. Following this 
procedure we can construct the complete set of translation rules for the 
imaginary units $e_{m}$ and the barred operators $1\mid e_{m}$ 
(appendix A). In this paper we will 
use the notation of refs.~\cite{jos1,jos2,dav}: $L_m$ and $R_m$ will represent 
the matrix 
counterpart of the octonionic operators $e_m$ and $1\mid e_m$,
\bel{j1}
L_{m}~\lrw ~e_{m} \q \mbox{and} \q R_{m} ~\lrw ~1\mid e_{m} \q .
\ee

At first glance it seems that our translation doesn't work. If we extract 
the matrices corresponding to $e_{1}$, $e_{2}$ and 
$e_{3}$, namely,
\[
L_{1}= \left(
\begin{array}{cccccccc}
0 & $-1$ & 0 & 0 & 0 & 0 & 0 &0\\
1 & 0 & 0 & 0 & 0 & 0 & 0 &0\\
0 & 0 & 0 & $-1$ & 0 & 0 & 0 &0\\
0 & 0 & 1 & 0 & 0 & 0 & 0 &0\\
0 & 0 & 0 & 0 & 0 & $-1$ &0 &0\\
0 & 0 & 0 & 0 & 1 & 0 & 0 &0\\
0 & 0 & 0 & 0 & 0 & 0 & 0 &1\\
0 & 0 & 0 & 0 & 0 & 0 & $-1$ &0
\end{array}
\right) \q , \q 
L_{2} = \left(
\begin{array}{cccccccc}
0 & 0 & $-1$ & 0 & 0 & 0 & 0 &0\\
0 & 0 & 0 & 1 & 0 & 0 & 0 &0\\
1 & 0 & 0 & 0 & 0 & 0 & 0 &0\\
0 & $-1$ & 0 & 0 & 0 & 0 & 0 &0\\
0 & 0 & 0 & 0 & 0 & 0 &$-1$ &0\\
0 & 0 & 0 & 0 & 0 & 0 & 0 & $-1$\\
0 & 0 & 0 & 0 & 1 & 0 & 0 & 0\\
0 & 0 & 0 & 0 & 0 & 1 & 0 &0
\end{array}
\right) \q , \q 
L_{3} =  \left(
\begin{array}{cccccccc}
0 & 0 & 0 & $-1$ & 0 & 0 & 0 &0\\
0 & 0 & $-1$ & 0 & 0 & 0 & 0 &0\\
0 & 1 & 0 & 0 & 0 & 0 & 0 &0\\
1 & 0 & 0 & 0 & 0 & 0 & 0 &0\\
0 & 0 & 0 & 0 & 0 & 0 &0 & $-1$\\
0 & 0 & 0 & 0 & 0 & 0 & 1 &0\\
0 & 0 & 0 & 0 & 0 & $-1$ & 0 &0\\
0 & 0 & 0 & 0 & 1 & 0 & 0 &0
\end{array}
\right) \q ,
\]
we find 
\be
L_{1}L_{2} \neq L_{3} \q .
\ee
In obvious contrast with the octonionic relation
\be
e_1 e_2 = e_3 \q .
\ee
This bluff is soon explained. In deducing our translation rules, we 
understand octonions as operators, and so they must be applied to a 
certain octonionic function, $\vpg$, and {\tt not} upon another ``operator''. 
So the octonionic relation
\bml
\be
e_3 \vpg~~[~=(e_1 e_2)\vpg~]  
\ee
is translated by
\be
L_3 \vpg \q ,
\ee
\eml
whereas, 
\bml
\be
e_1 (e_2 \vpg) ~~[~\neq e_3 \vpg~] 
\ee
becomes
\be
L_1 L_2 \vpg ~~[~\neq L_3 \vpg~] \q .
\ee
\eml
We have to differentiate between two kinds of multiplication, 
``~$\cdot$~''
and ``~$\times $~''. At the level of octonions, one has 
\be
e_1 \cdot e_2 = e_3 \q ,
\ee
but at level of octonionic operators
\be
e_1 \times e_2 \neq e_3 \q .
\ee
For $e_m$ and $1\mid e_m$, we have  simple ``$~\times ~$''-multiplication 
rules. In fact, utilizing the associator properties we find
\bml
\beal{pp}
e_m(e_n\vpg) & ~=~ & (e_m e_n)\vpg + (e_m\vpg)e_n - e_m (\vpg e_n) \q ,\\ 
(\vpg e_m) e_n & ~=~ & \vpg (e_m e_n) - (e_m\vpg)e_n + e_m (\vpg e_n) \q .
\eea
\eml
Thus, 
\bml
\bea
e_m ~\times ~ e_n & ~\equiv~ & 
-\dg_{mn} + \eg_{mnp} e_p + e_m~)~e_n - e_m ~(~e_n \q ,\\ 
~[~1\mid e_n~] ~\times ~ [~1\mid e_m~] & ~\equiv~ & 
-\dg_{mn} + \eg_{mnp} e_p - e_m~)~e_n + e_m ~(~e_n \q .
\eea
\eml
The previous relation can be soon rewritten in matrix form as follows
\cite{jos1}
\bml
\bea
L_m L_n & ~\equiv~ & 
-\dg_{mn} + \eg_{mnp} L_p + [R_n, \; L_m] \q ,\\ 
R_n R_m & ~\equiv~ & 
-\dg_{mn} + \eg_{mnp} R_p + [L_m, \; R_n] \q .
\eea
\eml
Introducing a new matrix multiplication, ``~$\circ$~'', related to the 
standard matrix multiplication (row by column) by
\bel{new}
L_m \circ L_n  \equiv L_m L_n + [R_n , \; L_m ] \q ,
\ee
we can quickly reproduce the nonassociative octonionic algebra
\be
L_m \circ L_n = -\dg_{mn} + \eg_{mnp} L_p \q .
\ee

Working with left/right barred operators we show how the nonassociativity is 
inherent in our representation. Such operators enable us to reproduce the 
octonions nonassociativity by the matrix algebra. Consider for example
\be
[~e_{3}~)~e_{1}~]~ \vpg ~ \equiv ~ (e_3 \vpg)e_1 
~=~  e_2 \vpg_0 - e_3 \vpg_1 + \vpg_2 - e_1 
\vpg_3 - e_6 \vpg_4
- e_7 \vpg_5 + e_4 \vpg_6 +e_5 \vpg_7 \q .
\ee
This equation will be translated into
\be
\left(
\begin{array}{cccccccc}
0 & 0 & 1 & 0 & 0 & 0 & 0 &0\\
0 & 0 & 0 &$-1$ & 0 & 0 & 0 &0\\
1 & 0 & 0 & 0 & 0 & 0 & 0 &0\\
0 &$-1$ & 0 & 0 & 0 & 0 & 0 &0\\
0 & 0 & 0 & 0 & 0 & 0 &1 &0\\
0 & 0 & 0 & 0 & 0 & 0 & 0 &1\\
0 & 0 & 0 & 0 & $-1$ & 0 & 0 &0\\
0 & 0 & 0 & 0 & 0 & $-1$ & 0 &0
\end{array}
\right)
\left( \begin{array}{c}
\vpg_0\\
\vpg_1\\
\vpg_2\\
\vpg_3\\
\vpg_4\\
\vpg_5\\
\vpg_6\\
\vpg_7
\end{array}
\right) = 
\left( \begin{array}{c}
\vpg_2\\
$-$\vpg_3 \\
\vpg_0\\
$-$\vpg_1\\
\vpg_6 \\
\vpg_7\\
$-$\vpg_4\\
$-$\vpg_5
\end{array} \right) 
\q .
\ee
Whereas,
\be
[~e_{3}~(~e_{1}~]~ \vpg ~ \equiv ~ e_3 (\vpg e_1)
~=~  e_2 \vpg_0 - e_3 \vpg_1 + \vpg_2  - e_1 
\vpg_3   + e_6 \vpg_4 + e_7 \vpg_5 - e_4 \vpg_6 - e_5 \vpg_7 \q ,     
\ee
will become 
\be
\left(
\begin{array}{cccccccc}
0 & 0 & 1 & 0 & 0 & 0 & 0 &0\\
0 & 0 & 0 &$-1$ & 0 & 0 & 0 &0\\
1 & 0 & 0 & 0 & 0 & 0 & 0 &0\\
0 &$-1$ & 0 & 0 & 0 & 0 & 0 &0\\
0 & 0 & 0 & 0 & 0 & 0 & $-1$ &0\\
0 & 0 & 0 & 0 & 0 & 0 & 0 & $-1$\\
0 & 0 & 0 & 0 & 1 & 0 & 0 &0\\
0 & 0 & 0 & 0 & 0 & 1 & 0 &0
\end{array}
\right)
\left( \begin{array}{c}
\vpg_0\\
\vpg_1\\
\vpg_2\\
\vpg_3\\
\vpg_4\\
\vpg_5\\
\vpg_6\\
\vpg_7
\end{array}
\right) = 
\left( \begin{array}{c}
\vpg_2\\
$-$\vpg_3 \\
\vpg_0\\
$-$\vpg_1\\
$-$\vpg_6 \\
$-$\vpg_7\\
\vpg_4\\
\vpg_5
\end{array} \right)  
\q . 
\ee
The nonassociativity is then reproduced since left and right barred 
operators, like
\[ e_{3}~)~e_{1} \q \mbox{and} \q  e_{3}~(~e_{1} \]
are represented by different matrices. 
The complete set of translation rules for 
left/right-barred operators is given in appendix A.

The matrix representation for left/right barred operators can be quickly 
obtained by suitable multiplications of the matrices $L_{m}$ and $R_{m}$. 
Let us clear up our assertion. By direct calculations we can extract the 
matrices which correspond to the operators
\[ e_m~)~e_n \qq \mbox{and} \qq e_{m}~(~e_n \qq ,\]
which we call, respectively,
\[ M^{L}_{mn} \qq \mbox{and} \qq M^{R}_{mn}  \q .\]
Our left/right barred operators can be represented by an ordered action of 
the operators $e_{m}$ and $1\mid e_{m}$, and so we can related the matrices 
$M^{L}_{mn}$ and  $M^{R}_{mn}$ to the matrices 
$L_m$ and $R_m$:
\bml
\bea
M^{L}_{mn} & ~\equiv ~& R_n L_m \q ,\\
M^{R}_{mn} & ~\equiv ~& L_m R_n \q .
\eea
\eml

The previous discussions concerning the octonions nonassociativity and the 
isomorphism between $\glr$ and barred octonions, can be 
now, elegantly, presented as follows.\\
{\tt 1 - Matrix representation for octonions nonassociativity.} 
\bea
M^{L}_{mn} ~\neq ~ M^{R}_{mn} \qq 
[ ~R_n L_m  \neq L_m R_n ~~ \mbox{{\fs for $m\neq n$}} ] \q .
\eea
{\tt 2 - Isomorphism between} \bfm{\glr} {\tt and barred octonions.}\\
If we rewrite our 106 barred operators by real matrices:
\bc
\bt{lr}
$1, \; L_{m}, \; R_{m}$ & {\fs ~~~~~(15 matrices)} ,\\
$M\equiv L_m R_m =R_m L_m $ &{\fs (7)} ,\\
$M^{L}_{mn}\equiv R_n L_{m}$ ~~~~{\fs $(m\neq n)$} & {\fs (42)} ,\\
$M^{R}_{mn}\equiv L_n R_{m}$ ~~~~{\fs $(m\neq n)$} & {\fs (42)} ,\\
{\fs $(m, \; n =1, ..., 7) \q ;$} & 
\et
\ec
we have two different basis for $\glr$:
\bc
\bt{cl}
{\fs (1)} & ~~~~~$1 \; , ~ L_{m} \; , ~ R_{m} \; , ~ R_n L_{m} \q ,$\\ 
 & \\
{\fs (2)} & ~~~~~$1 \; , ~ L_{m} \; , ~ R_{m} \; , ~ L_{m} R_n \q .$ 
\et
\ec

We now remark some difficulties deriving from the octonions nonassociativity. 
When we translate from barred octonions
to $8 \times 8$ real matrices there is no problem. For example, in the 
octonionic equation
\bel{ww}
e_{4} \{[(e_{6}\vpg ) e_{1}]e_{5}\} \q , 
\ee
we quickly recognize the following left-barred operators,
\[
e_{4}~(~e_{5} \q \mbox{and} \q e_{6}~)~e_{1}  \q .
\]
We can translate eq.~(\ref{ww}) into
\be
M^{L}_{45}~M^{L}_{61}~\vpg \q .
\ee
In going from $8\times 8$ real matrices to octonions we should 
be careful in ordering. For example,
\bel{mm}
A B ~\vpg ~~~
\ee
can be understood as
\bml
\bel{mm1}
(AB) \vpg \q ,
\ee
or
\bel{mm2}
A (B \vpg) \q .
\ee
\eml
The first choice is related to the  ``$~\times ~$'' multiplication 
(different from the standard octonionic multiplication). In order to avoid 
confusion we  
translate eq.~(\ref{mm}) by eq.~(\ref{mm2}).  In general 
\be
ABC \ldots Z \vpg \equiv A(B(C \ldots (Z\vpg) \ldots )) \q .
\ee

\subsection*{IV-b. Relation between barred operators and $4\times 4$ complex 
matrices}

Some complex groups play a critical role
in physics. No one can deny the importance
of $U(1, \co)$ or $SU(2, \co)$. In relativistic
quantum mechanics, $\glc$ is essential in writing the 
Dirac equation. Having $\glr$, we should be able
to extract its subgroup $\glc$. So, we can 
translate the famous Dirac-gamma matrices and 
write down a one-component octonionic Dirac equation~\cite{odir}.

Let us show how we can isolate our 32 basis of $\glc$:\\
Working with the symplectic decomposition of octonions
\be
\psi = \baq{\psi_{1}}{\psi_{2}}{\psi_{3}}{\psi_{4}} ~ \lrw ~
\psi_{1} + e_{2} \psi_{2} + e_{4} \psi_{3} + e_{6} \psi_{4} \qq 
[~\psi_{1,  ..., 4} \in \bfm{\cal C}(1, \; e_{1})~] \q .
\ee
we analyse the action of 
left-barred operators on our octonionic wave functions $\psi$. For example, 
we find 
\bean
~[~1\mid e_{1}~]~\psi ~\equiv ~ & \psi e_{1} & ~=~ \psi_{1} + e_{2} (e_{1}\psi_{2}) + 
e_{4} (e_{1}\psi_{3}) + e_{6} (e_{1}\psi_{4}) \q ,\\
 & e_{2}\psi & ~=~  -\psi_{2} + e_{2} \psi_{1} -
e_{4} \psi_{4}^{*} + e_{6} \psi_{3}^{*} \q ,\\
~[~e_{3}~)~e_{1}~]~\psi ~\equiv ~ & (e_{3}\psi) e_{1} & ~=~ 
\psi_{2} + e_{2} \psi_{1} +
e_{4} \psi_{4}^{*} - e_{6} \psi_{3}^{*} \q .
\eean

Following the same methodology of the previous section,
we can immediately note a correspondence between 
the complex matrix $i \openone_{4\times 4}$ and the octonionic   
barred operator $1\mid e_{1}$
\be
\bamq{i}{0}{0}{0} 0 & i & 0 & 0\\ 0 & 0 & i & 0\\ 0 & 0 & 0 & i \eam
~ \lrw ~1\mid e_{1} \q .
\ee

The translation doesn't work for all barred operators. Let us show it, 
explicitly. For example, we cannot find a $4\times 4$ complex matrix which, 
acting on
\[
\baq{\psi_{1}}{\psi_{2}}{\psi_{3}}{\psi_{4}} \q , 
\]
gives the column vector
\[
\left( \begin{array}{c}
$-$\psi_{2} \\ \psi_{1} \\ $-$\psi_{4}^{*} \\ \psi_{3}^{*}
\end{array} \right) \qq \mbox{or} \qq 
\left( \begin{array}{c} 
\psi_{2} \\ \psi_{1} \\ \psi_{4}^{*} \\ $-$\psi_{3}^{*}
\end{array} \right) \q ,
\]
and so we have not the possibility to relate 
\[ e_{2} \qq \mbox{or} \qq  e_{3}~)~e_{1} \]
with a complex matrix. Nevertheless, a combined action of such operators
gives us
\[
e_{2}\psi + (e_{3}\psi)e_{1} = 2 e_{2}\psi_{1} \q ,
\]
and it allows us to represent the octonionic barred operator 
\bml
\be 
e_{2} \; + \; e_{3}~)~e_{1} \q , 
\ee
by the $4\times 4$ complex matrix
\bel{ct}
\bamq{0}{0}{0}{0} 2 & 0 & 0 & 0\\ 0 & 0 & 0 & 0\\ 0 & 0 & 0 & 0 \eam \q .
\ee
\eml
Following this procedure we can represent a generic $4\times 4$ complex 
matrix by octonionic barred operators. The explicit correspondence rules are 
given in appendix B.

We conclude our discussion concerning the relation between barred operators 
and $4 \times 4$ complex matrices, noting that the 32 basis elements of 
$GL(4,\co)$ can be directly extract from the 64 generators of 
$GL(8,\rea)$. It is well known that any complex matrix can be rewritten as 
a real matrix by the following isomorphism
\[
1~\lrw ~ \openone_{2\times 2} \qq \mbox{and} \qq 
i~\lrw ~ -i\sigma_2 \q .
\]
The situation at the lowest order is
\bean 
GL(2,\rea) & ~~~~~\mbox{generators :}~~~~~ & 
\openone_{2\times 2} ~, ~ \sigma_1 ~, ~-i\sigma_2 ~, ~ \sigma_3 \q ;\\
GL(1,\co)~ & ~~~~~\mbox{isomorphic :}~~~~~ & 
\openone_{2\times 2} ~, ~-i\sigma_2  \q .
\eean
In a similar way (choosing appropriate combinations of left-barred 
octonionic operators, in which only $\pm \openone_{2\times 2}$ and 
$\pm i\sigma_2$ appear) we can extract from $GL(8,\rea)$ the 32 basis 
elements of $GL(4,\co)$. For further details see appendix B. 

\subsection*{IV-c. Octonionic representations of the 4-dimensional Clifford 
algebra}

We show explicitly two octonionic representations for the Dirac 
gamma-matrices~\cite{itz}:
\bc
1-{\tt Dirac representation,}
\ec
\bml
\beal{odgm1}
\cg^{0} & = & \frac{1}{3} -\frac{2}{3} \sum_{m=1}^{3} e_{m}\mid e_{m} +
\frac{1}{3} \sum_{n=4}^{7} e_{n} \mid e_{n} \q ,\\
\cg^{1} & = & -\frac{2}{3} e_{6} -\frac{1}{3}\mid e_{6} + e_{5}~)~e_{3} - 
e_{3}~)~e_{5}  - \frac{1}{3} \sum_{p, \; s =1}^{7} \eg_{ps6} e_{p}~)~e_{s} 
\q ,\\
\cg^{2} & = & -\frac{2}{3} e_{7} -\frac{1}{3}\mid e_{7} + e_{3}~)~e_{4} - 
e_{4}~)~e_{3}  - \frac{1}{3} \sum_{p, \; s =1}^{7} \eg_{ps7} e_{p}~)~e_{s} 
\q ,\\
\cg^{3} & = & -\frac{2}{3} e_{4} -\frac{1}{3}\mid e_{4} + e_{7}~)~e_{3} - 
e_{3}~)~e_{7}  - \frac{1}{3} \sum_{p, \; s =1}^{7} \eg_{ps4} e_{p}~)~e_{s} 
\q ;
\eea
\eml
\bc
2-{\tt Majorana representation,}
\ec
\bml
\beal{odgm2}
\cg^{0} & = & 
\frac{1}{3}  e_7 - 
\frac{1}{3} \mid e_7 
+ e_3~)~e_4 - e_5~)~e_2 + e_6~)~e_1
- \frac{1}{3} \sum_{p, \; s=1}^{7} \eg_{ps7} e_p~)~e_s \q ,\\
\cg^{1} & = & \frac{2}{3} e_{1}  + \frac{1}{3}\mid e_{1} 
+ e_{5}~)~e_{4} 
 - e_{4}~)~e_{5}  + \frac{1}{3} \sum_{p, \; s =1}^{7} 
\eg_{ps1} e_{p}~)~e_{s} 
\q ,\\
\cg^{2} & = & \frac{2}{3} e_{7} +\frac{1}{3}\mid e_{7} 
+ e_{4}~)~e_{3} -  
e_{3}~)~e_{4}  + \frac{1}{3} \sum_{p, \; s =1}^{7} \eg_{ps7} e_{p}~)~e_{s} 
\q ,\\
\cg^{3} & = & \frac{2}{3} e_{3} +\frac{1}{3}\mid e_{3} + 
e_{7}~)~e_{4} - 
e_{4}~)~e_{7}  + \frac{1}{3} \sum_{p, \; s =1}^{7} \eg_{ps3} e_{p}~)~e_{s} 
\q .
\eea
\eml

\con

The modern notion of symmetry in physics heavily depends upon using the 
associative Lie groups. So, at first glance, it seems that octonions have 
not any relation with our physical world. Having a nonassociative algebra 
needs special care. In this work, we introduced a ``trick'' which allowed us 
to manipulate octonions without useless efforts, namely 
{\tt barred octonions}. 

This paper aimed to give a clear exposition of the potentiality of 
{\tt barred numbers}. Their possible applications could occur in different 
fields, like group theory, quantum mechanics, nuclear physics. We preferred 
in our work to focus our attention on the mathematical subject. Physical 
applications are investigated elsewhere~\cite{deleos,odir,oqm}.

We summarize the more important results found in previous sections:
\bc
{\tt M - Mathematical Contents :}
\ec

{\tt M1} - The introduction of barred operators (natural objects if one works 
with noncommutative numbers) facilitate our job and enable us to formulate 
a ``friendly'' connection between $8 \times 8$ real matrices and octonions;

{\tt M2} - The nonassociativity is reproduced by left/right barred 
operators. We consider these operators the natural extension of barred 
quaternions, recently introduced in literature~\cite{qua2,deleos};

{\tt M3} - We tried to investigate the properties of our barred 
numbers and studied their special characteristics in order to use them in a 
proper way. After having established their isomorphism to $\glr$, life became 
easier;

{\tt M4} -  The connection between $\glr$ and barred octonions gives us 
the possibility to extract the octonionic generators corresponding to the 
complex subgroup $\glc$. This step represents the main tool to manipulate 
octonions in quantum mechanics;

{\tt M5} - To the best of our knowledge, for the first time, an octonionic 
representation for the 4-dimensional Clifford algebra, appears in print.

\bc
{\tt I - Further Investigations :}
\ec

We conclude with a listing of open questions for future investigations, 
whose study lead to further insights.

{\tt I1} - How may we complete the translation? Note that translation, as 
presented in this paper, works for $4n\times 4n$ matrices. What about 
odd-dimensional matrices?

{\tt I2} - From the translation rules we can extract the multiplication 
rules for generic octonionic barred operators. This will allow us to work 
directly with octonions without translations. 

{\tt I3} - Inspired from eq.~(\ref{new}), we could look for a more 
convenient way to express the new nonassociative multiplication 
(for example we can try to modify the standard 
multiplication rule: row by column);

{\tt I4} - A last interesting research topic could be to generalize the group 
theoretical structure by our barred octonionic operators.

Many of the problems on this list deal with technical details although the 
answers to some will be important for further development of the subject.

We hope that the work presented in this paper, demonstrates that octonions 
may constitute a coherent and well-defined branch of theoretical research. 
We are convinced that octonions represent largely uncharted and potentially 
very interesting terrain for theoretical investigations.

\ack

The authors would like to thank P.~Rotelli for invaluable conversations and 
suggestions. One of us (KAK) gratefully 
acknowledges the warm hospitality of the Lecce Physics Department, where this 
paper was prepared.  The authors also thank the referee for improvements to 
this work.

\section*{Appendix A\\ Octonionic Representation of
GL(8,${\cal R}$)}

In this appendix we give the translation rules between octonionic 
left-right barred operators and $8\times 8$ real matrices. In order to 
simplify our presentation we introduce the following notation:
\bml
\bea
\{~a, \; b, \; c, \; d~\}_{(1)} ~ \equiv ~\bamq{a}{0}{0}{0} 0 & b & 0 & 0\\
0 & 0 & c & 0\\ 0 & 0 & 0 & d \eam   & \q , \q &
\{~a, \; b, \; c, \; d~\}_{(2)} ~ \equiv ~\bamq{0}{a}{0}{0} b & 0 & 0 & 0\\
0 & 0 & 0 & c\\ 0 & 0 & d & 0 \eam \q ,\\
\{~a, \; b, \; c, \; d~\}_{(3)} ~ \equiv ~\bamq{0}{0}{a}{0} 0 & 0 & 0 & b\\
c & 0 & 0 & 0\\ 0 & d & 0 & 0 \eam   & \q , \q &
\{~a, \; b, \; c, \; d~\}_{(4)} ~ \equiv ~\bamq{0}{0}{0}{a} 0 & 0 & b & 0\\
0 & c & 0 & 0\\ d & 0 & 0 & 0 \eam \q ,
\eea
\eml
where $a, \; b, \; c, \; d$ and $0$ represent $2\times 2$ real matrices.

From now on, with $\sg_{1}$, $\sg_{2}$, $\sg_{3}$ we represent the 
standard Pauli matrices:
\be
\sg_{1} = \bamd{0}{1}{1}{0} \q , \q  
\sg_{2} = \bamd{0}{-i}{i}{0} \q , \q  
\sg_{3} = \bamd{1}{0}{0}{-1} \q .  
\ee

The only necessary translation rules that we need to know 
explicitly are the following
\bc
\bt{rcrrrrcrcrrrrc} 
 & & & & & & & & & & & & & \\ 
\bfm{e_{1}} & $~\lrw~\{$ & $-i\sg_{2}$, & ~$-i\sg_{2}$, & ~$-i\sg_{2}$, & 
~$i\sg_{2} ~\}_{(1)}$
&~~~,~~~~& 
\bfm{1\mid e_{1}} & $~\lrw~\{$ & $-i\sg_{2}$, & ~$i\sg_{2}$, & ~$i\sg_{2}$, & 
~$-i\sg_{2} ~\}_{(1)}$
&~~~,~~~~\\
\bfm{e_{2}} & $~\lrw~\{$ & $ -\sg_{3}$, & ~$\sg_{3}$, & ~$-\openone$, & 
~$\openone ~\}_{(2)}$
&~~~,~~~~& 
\bfm{1\mid e_{2}} & $~\lrw~\{$ & $ -\openone$, & ~$\openone$, & ~$\openone$, & 
~$-\openone ~\}_{(2)}$
&~~~,~~~~\\
\bfm{e_{3}} & $~\lrw~\{$ & $ -\sg_{1}$, & ~$\sg_{1}$, & ~$-i\sg_{2}$, & 
~$-i\sg_{2} ~\}_{(2)}$
&~~~,~~~~& 
\bfm{1\mid e_{3}} & $~\lrw~\{$ & $ -i\sg_{2}$, & ~$-i\sg_{2}$, & ~$i\sg_{2}$, & 
~$i\sg_{2} ~\}_{(2)}$
&~~~,~~~~\\
\bfm{e_{4}} & $~\lrw~\{$ & $ -\sg_{3}$, & ~$\openone$, & ~$\sg_{3}$, & 
~$-\openone ~\}_{(3)}$
&~~~,~~~~& 
\bfm{1\mid e_{4}} & $~\lrw~\{$ & $ -\openone$, & ~$-\openone$, & ~$\openone$, & 
~$\openone ~\}_{(3)}$
&~~~,~~~~\\
\bfm{e_{5}} & $~\lrw~\{$ & $ -\sg_{1}$, & ~$i\sg_{2}$, & ~$\sg_{1}$, & 
~$i\sg_{2} ~\}_{(3)}$
&~~~,~~~~& 
\bfm{1\mid e_{5}} & $~\lrw~\{$ & $ -i\sg_{2}$, & ~$-i\sg_{2}$, & ~$-i\sg_{2}$, & 
~$-i\sg_{2} ~\}_{(3)}$
&~~~,~~~~\\
\bfm{e_{6}} & $~\lrw~\{$ & $ -\openone$, & ~$-\sg_{3}$, & ~$\sg_{3}$, & 
~$\openone ~\}_{(4)}$
&~~~,~~~~& 
\bfm{1\mid e_{6}} & $~\lrw~\{$ & $ -\sg_{3}$, & ~$\sg_{3}$, & ~$-\sg_{3}$, & 
~$\sg_{3} ~\}_{(4)}$
&~~~,~~~~\\
\bfm{e_{7}} & $~\lrw~\{$ & $ -i\sg_{2}$, & ~$-\sg_{1}$, & ~$\sg_{1}$, & 
~$-i\sg_{2} ~\}_{(4)}$
&~~~,~~~~& 
\bfm{1\mid e_{7}} & $~\lrw~\{$ & $ -\sg_{1}$, & ~$\sg_{1}$, & ~$-\sg_{1}$, & 
~$\sg_{1} ~\}_{(4)}$
&~~~.~~~~\\
\et
\ec
The remaining rules can be easily constructed remembering that
\bean
\bfm{e_m}             & ~~\lrw ~~& L_m \q ,\\
\bfm{1 \mid e_m}      & ~~\lrw ~~& R_m \q ,\\
\bfm{e_m \mid e_m}    & ~~\lrw ~~& M^{L}_{mm} ~\equiv ~ R_m L_m \q ,\\
                &      & M^{R}_{mm}  ~\equiv ~ L_m R_m \q ,\\  
\bfm{e_m ~)~ e_n}     & ~~\lrw ~~& M^{L}_{mn} ~\equiv ~  R_n L_m \q ,\\ 
\bfm{e_m ~( ~e_n}     & ~~\lrw ~~ & M^{R}_{mn}  ~\equiv ~ L_m R_n \q . 
\eean
For example,
\bean
\bfm{e_1 \mid e_1} ~~\lrw~~ 
\left( \begin{array}{cccc}
-i \sg_2 & 0 & 0 & 0 \\
0 & -i \sg_2 & 0 & 0\\
0 & 0 & -i \sg_2 & 0 \\
0 & 0 & 0 & i \sg_2
 \end{array} \right)
\left( \begin{array}{cccc}
-i \sg_2 & 0 & 0 & 0 \\
0 & i \sg_2 & 0 & 0\\
0 & 0 & i \sg_2 & 0 \\
0 & 0 & 0 & -i \sg_2
 \end{array} \right)
 &~=~& \{ \;  -\openone, \; \openone, \; \openone, \; \openone \; \}_{(1)} \q ,
\eean
\bean 
\bfm{e_3 ~)~ e_1} ~~\lrw ~~ 
\left( \begin{array}{cccc}
-i \sg_2 & 0 & 0 & 0 \\
0 &i \sg_2 & 0 & 0\\
0 & 0 & i \sg_2 & 0 \\
0 & 0 & 0 & - i \sg_2 
 \end{array} \right)
\left( \begin{array}{cccc}
0 & -\sg_1 & 0 & 0 \\
\sg_1 & 0 & 0 & 0\\
0 & 0 & 0 & -i \sg_2 \\
0 & 0 & - i \sg_2 & 0
 \end{array} \right)
~=~ \{ \;  \sg_3, \; \sg_3 , \; \openone , \; - \openone \; \}_{(2)} \q ,
\eean
and
\bean \bfm{e_3  ~( ~e_1}  ~~\lrw ~~ 
\left( \begin{array}{cccc}
0 & -\sg_1 & 0 & 0 \\
\sg_1 & 0 & 0 & 0\\
0 & 0 & 0 & -i \sg_2 \\
0 & 0 & - i \sg_2 & 0
 \end{array} \right)
\left( \begin{array}{cccc}
-i \sg_2 & 0 ^ 0 & 0 \\
0 &i \sg_2 & 0 & 0\\
0 & 0 & i \sg_2 & 0 \\
0 & 0 & 0 & - i \sg_2 
 \end{array} \right)
&~=~& \{ \; \sg_3, \; \sg_3 , \; - \openone , \; \openone \}_{(2)} \q .
\eean
Following this procedure any matrix representation of
right/left barred operators can be obtained. Using 
Mathematica~\cite{math},  we have proved the linear independence of
the 64 elements which represent the most general octonionic operator  
\[
{\cal O}_{0}+\sum_{m=1}^{7} {\cal O}_{m}~)~e_{m} \q .
\]
So our barred operators form a complete basis for any $8\times 8$ real 
matrix and this establishes the isomorphism between $\glr$ and barred 
octonions.

We conclude this appendix giving a compact notation for the 64 
left-barred operators (a similar trick works for the right ones). 

For $m,n = 1 \ldots 7 ~(m \neq n)$  and $\alpha , \beta = 1 
\dots 7$ (labels of the rows and columns of the corresponding 
matrix $X$), we have 
\bml
\bea
\bfm{e_m ~)~ e_n} + \bfm{e_n ~) ~e_m}  
~~\lrw~~ X_{\alpha \beta} =
\left\{ 
\begin{array}{cc} 
$-2$ & ~~\alpha, \beta ~=~ m,n~;~ n,m~. \\ 
0    & \mbox{otherwise .}
\end{array}
\right.
\eea
\bea
\bfm{e_m} + \bfm{1 \mid e_m} 
~~\lrw ~~X_{\alpha \beta} =
\left\{ 
\begin{array}{cc}
$-2$    & ~~\alpha, \beta ~=~ 0,m~.  \\
$+2$    & ~~\alpha, \beta ~=~ m,0~.  \\
0       & \mbox{otherwise .}
\end{array}
\right.
\eea
\eml

For the minus combination, after introducing the index p
defined by $e_m e_n = \epsilon_{m n p}$ ($m \neq n$), we have 
the following rules 
\bml
\bea
\bfm{e_m ~)~ e_n} - \bfm{e_n~ )~ e_m }
~~\lrw ~~X_{\alpha \beta} =
\left\{ 
\begin{array}{cc} 
$2$\epsilon_{a b p}  & ~~\alpha, \beta ~=~ a, b~ (\neq m, n)~. \\ 
$2$ & ~~\alpha, \beta ~= ~0,p ~; ~p,0~. \\
0    & \mbox{otherwise .}
\end{array}
\right.
\eea
\bea
\bfm{e_m} - \bfm{1 \mid e_m}  
~~\lrw ~~X_{\alpha \beta} =
\left\{ 
\begin{array}{cc}
$-2$ \epsilon_{a b m} 
     & ~\alpha, \beta ~= ~a,b~. \\ 
0    & \mbox{otherwise .}
\end{array}
\right.
\eea
\eml

\section*{Appendix B\\ Octonionic Representation of GL(4,${\cal C}$)}
We give the action of barred operators on octonionic functions
\[ \psi = \psi_{1} +e_{2}\psi_{2}+e_{4}\psi_{3}+e_{6}\psi_{4} \qq
[~\psi_{1, ...,4} \in \co (1, \; e_{1})~] \q .
\]
In the following we will use the notation
\[
e_{2} ~\raw ~ \{ -\psi_{2}, \; \psi_{1}, \;
-\psi_{4}^{*}, \; \psi_{3}^{*} \} \q ,
\]
to indicate
\[ 
e_{2}\psi~=~ -\psi_{2}+e_{2}\psi_{1}-e_{4}\psi_{4}^{*}+e_{6}\psi_{3}^{*} \q .
\]

As occurred in the previous appendix we need to know only the action of the 
barred operators \bfm{e_{m}} and \bfm{1\mid e_{m}}
\bc
\bt{rcrrrrcrcrrrrc}
\bfm{e_{1}} & $~\raw~\{$ & $e_{1}\psi_{1}$, & ~$-e_{1}\psi_{2}$, & 
~$-e_{1}\psi_{3}$, & 
~$-e_{1}\psi_{4} ~\}$
&~~,~~~& 
\bfm{1\mid e_{1}} & $~\raw~\{$ & $e_{1}\psi_{1}$, & ~$e_{1}\psi_{2}$, & 
~$e_{1}\psi_{3}$, & 
~$e_{1}\psi_{4} ~\}$
&~~,~~~\\
\bfm{e_{2}} & $~\raw~\{$ & $ -\psi_{2}$, & ~$\psi_{1}$, 
& ~$-\psi_{4}^{*}$, & ~$\psi_{3}^{*} ~\}$
&~~,~~~& 
\bfm{1\mid e_{2}} & $~\raw~\{$ & $ -\psi_{2}^{*}$, & ~$\psi_{1}^{*}$, 
& ~$\psi_{4}^{*}$, & 
~$-\psi_{3}^{*} ~\}$
&~~,~~~\\
\bfm{e_{3}} & $~\raw~\{$ & $ -e_{1}\psi_{2}$, & ~$-e_{1}\psi_{1}$, 
& ~$-e_{1}\psi_{4}^{*}$, & 
~$e_{1}\psi_{3}^{*} ~\}$
&~~,~~~& 
\bfm{1\mid e_{3}} & $~\raw~\{$ & $ e_{1}\psi_{2}^{*}$, & ~$-e_{1}\psi_{1}^{*}$, 
& ~$e_{1}\psi_{4}^{*}$, & 
~$-e_{1}\psi_{3}^{*} ~\}$
&~~,~~~\\
\bfm{e_{4}} & $~\raw~\{$ & $ -\psi_{3}$, & ~$\psi_{4}^{*}$, 
& ~$\psi_{1}$, & ~$-\psi_{2}^{*} ~\}$
&~~,~~~& 
\bfm{1\mid e_{4}} & $~\raw~\{$ & $ -\psi_{3}^{*}$, & ~$-\psi_{4}^{*}$, 
& ~$\psi_{1}^{*}$, & 
~$\psi_{2}^{*} ~\}$
&~~,~~~\\
\bfm{e_{5}} & $~\raw~\{$ & $ -e_{1}\psi_{3}$, & ~$e_{1}\psi_{4}^{*}$, 
& ~$-e_{1}\psi_{1}$, & 
~$-e_{1}\psi_{2}^{*} ~\}$
&~~,~~~& 
\bfm{1\mid e_{5}} & $~\raw~\{$ & $ e_{1}\psi_{3}^{*}$, & ~$-e_{1}\psi_{4}^{*}$, 
&~$-e_{1}\psi_{1}^{*}$, & 
~$e_{1}\psi_{2}^{*} ~\}$
&~~,~~~\\
\bfm{e_{6}} & $~\raw~\{$ & $ -\psi_{4}$, & ~$-\psi_{3}^{*}$, &
 ~$\psi_{2}^{*}$, & 
~$\psi_{1} ~\}$
&~~,~~~&
\bfm{1\mid e_{6}} & $~\raw~\{$ & $ -\psi_{4}^{*}$, & ~$\psi_{3}^{*}$, 
& ~$-\psi_{2}^{*}$, & 
~$\psi_{1}^{*} ~\}$
&~~,~~~\\
\bfm{e_{7}} & $~\raw~\{$ & $ e_{1}\psi_{4}$, & ~$e_{1}\psi_{3}^{*}$, 
& ~$-e_{1}\psi_{2}^{*}$, & ~$e_{1}\psi_{1} ~\}$
&~~,~~~& 
\bfm{1\mid e_{7}} & $~\raw~\{$ & $ -e_{1}\psi_{4}^{*}$, 
& ~$-e_{1}\psi_{3}^{*}$, 
& ~$e_{1}\psi_{2}^{*}$, & 
~$e_{1}\psi_{1}^{*} ~\}$
&~~~.~~~
\et
\ec
From the previous correspondence rules we immediately obtain the others 
barred operators. We give, as example, the construction of the operator
\bfm{e_4~)~e_7}. We know that
\[ \bfm{e_{4}} ~\raw~ \{-\psi_{3}, \; \psi_{4}^{*}, \;  \psi_{1}, \; 
-\psi_{2} ^{*}\} \qq \mbox{and} \qq  
\bfm{1\mid e_{7}} ~\raw~ \{-e_{1}\psi_{4}^{*}, \;  
-e_{1}\psi_{3}^{*}, \;  e_{1}\psi_{2}^{*}, \; e_{1}\psi_{1}^{*} \}
\q .
\]
Combining these operators we find 
\[ \{-e_1(-\psi_{2}^*)^*, \; -e_1 \psi_{1}^{*}, \;  e_1 (\psi_{4}^*)^* , \; 
e_1 (-\psi_{3}) ^{*}\} \q ,
\]
and so
\[ \bfm{e_4 ~)~ e_{7}} ~\raw ~  
\{e_1\psi_{2}, \; -e_1 \psi_{1}^{*}, \;  e_1 \psi_{4} , \; 
- e_1 \psi_{3}^{*} \} \q . 
\]

As remarked at the end of subsection IV-b, we can extract the 32 basis 
elements of $GL(4,\co)$ directly by suitable combinations of 64 basis 
elements of $GL(8,\rea)$. We must choose the combination which have only 
$\openone_{2\times 2}$ and $-i\sigma_2$ as matrix elements. Nevertheless we 
must take care in manipulating our octonionic barred 
operators. If we wish to extract from $GL(8,\rea)$ the 32 elements 
which characterize $GL(4,\co)$ we need to change the 
octonionic basis of $GL(8,\rea)$. 
In fact, the natural choice for the symplectic octonionic representation
\bean 
\psi & ~=~ & (\vpg_{0} +e_{1}\vpg_{1}) + e_2 (\vpg_{2} +e_{1}\vpg_{3}) +
              e_4 (\vpg_{4} +e_{1}\vpg_{5}) +
              e_6 (\vpg_{6} +e_{1}\vpg_{7}) \q ,
\eean
requires the following real counterpart
\bean
\tilde{\vpg}  & ~=~ & \vpg_{0} +e_{1}\vpg_{1} + e_2 \vpg_{2} - e_{3}\vpg_{3} +
              e_4 \vpg_{4} - e_{5}\vpg_{5} +
              e_6 \vpg_{6} + e_{7}\vpg_{7} \q .
\eean  
whereas we used in subsection IV-a the following basis
\bean
\vpg  & ~=~ & \vpg_{0} +e_{1}\vpg_{1} + e_2 \vpg_{2} + e_{3}\vpg_{3} +
              e_4 \vpg_{4} + e_{5}\vpg_{5} +
              e_6 \vpg_{6} + e_{7}\vpg_{7} \q .
\eean  

The changes in the signs of $e_3\vpg_3$ and $e_5\vpg_5$ implies a 
modification in the generators of $GL(8,\rea)$. For example,
\bfm{e_2} and \bfm{e_3~)~e_1} now read
\[ \bfm{e_2}~\equiv ~\{-\openone, \; \openone, \; -\sigma_3, \; 
\sigma_3 \}_{(2)}
\qq \mbox{and} \qq \bfm{e_3~)~e_1}~\equiv ~\{\openone, \; \openone, \; 
\sigma_3, \; -\sigma_3 \}_{(2)} \q .
\]
i.e the change of basis induce the following modifications
\[ \openone \rightleftharpoons \sg_3 \q .\]
Their appropriate combination gives
\[ \frac{\bfm{e_2} + \bfm{e_3~)~e_1}}{2}~\equiv 
~\{0, \; \openone, \; 0, \; 0\}_{(2)}~~ 
\stackrel{complexifing}{\longrightarrow} 
~~ \left( \begin{array}{cccc}
0 & 0 & 0 & 0\\
1 & 0 & 0 & 0\\
0 & 0 & 0 & 0\\
0 & 0 & 0 & 0
\end{array}
\right) \q ,
\]
as required by eq.~(\ref{ct}). 

\bref
\bi{om1}
C.~Chevalley and R.~D.~Schafer, Proc.~Nat.~Acad.~Sci. {\bf 36}, 137 (1950).
\bi{om2}
J.~Tits, Proc.~Colloq.~Utrecht, 135 (1962).
\bi{om3}
H.~Freudenthal, Advances in Math.~I, 145 (1965).
\bi{om4}
R.~D.~Schafer,  {\it An introduction to Non-Associative Algebras} 
(Academic Press, New York, 1966).
\bi{qua2}
S.~De Leo and P.~Rotelli, \pxxa{92}{917}{94}.
\bi{rel}
S.~De Leo, {\it Quaternions and Special Relativity}, J.~Math.~Phys. 
(to be published).
\bi{alb2}
A.~A.~Albert, Ann.~Math. {\bf 48}, 495 (1947).
\bi{ham}
W.~R.~Hamilton, {\it Elements of Quaternions} (Chelsea Publishing Co., New 
York, 1969).
\bi{gra}
J.~T.~Graves, {\it Mathematical Papers} (1843).
\bi{cay}
A.~Cayley, Phil.~Mag.~(London) {\bf 26}, 210 (1845).
A.~Cayley, {\it Papers Collected Mathematical Papers} (Cambridge 1889). 
\bi{fro}
G.~Frobenius, J.~Reine Angew.~Math. {\bf 84}, 59 (1878).
\bi{hur}
A.~Hurwitz, Nachr.~Gesell.~Wiss.~G\"ottingeg, Math.~Phys.~Kl., 309.\\
A.~Hurwitz, {\it Mathematische Werke Band II, Zahlentheorie, Algebra und 
Geometrie}, pag.~565 (Birkha\"user, Basel, 1933).
\bi{bot}
R.~Bott and J.~Milnor, Bull.~Amer.~Math.~Soc. {\bf 64}, 87 (1958). 
\bi{ker}
M.~Kervaire, Proc.~Nat.~Acad.~Sci. {\bf 44}, 280 (1958).
\bi{oku}
S.~Okubo, {\it Introduction to Octonion and Other Non-Associative Algebras 
in Physics} (Cambridge University Press, Cambridge, to be published).
\bi{gur1}
F.~G\"ursey, {\it Symmetries in Physics (1600-1980): Proc.~of the 1st 
International Meeting on the History of Scientific Ideas}, Seminari 
d'~Hist\`oria de les Ci\`ences, Barcelona, Spain, 1987, p.~557.
\bi{adl}
S.~L.~Adler, {\it Quaternionic Quantum Mechanics and Quantum Fields} 
(Oxford, New York, 1995).
\bi{dir2}
P.~Rotelli, \mxb{4}{933}{89}.
\bi{deleos}
S.~De Leo and P.~Rotelli, \pxf{45}{575}{92}; \nxd{B110}{33}{95}; 
\ixa{10}{4359}{95}; Mod.~Phys.~Lett.~A {\bf 11}, 357 (1996).\\
S.~De Leo and P.~Rotelli, {\it Quaternionic Electroweak Theory}, 
(to be published in J.~Phys.~G).\\
S.~De Leo, \pxxa{94}{11}{95}; {\it Quaternions for GUTs}, 
Int.~J.~Theor.~Phys. (submitted).
\bi{mor2}
K.~Morita, \pxxa{67}{1860}{81}; \xxx{68}{2159}{82}; \xxx{70}{1648}{83}; 
\xxx{72}{1056}{84}; \xxx{73}{999}{84}; \xxx{75}{220}{85}; \xxx{90}{219}{93}. 
\bi{dir4}
S.~De Leo, {\it One-component Dirac Equation}, Int.~J.~Mod.~Phys.~A 
(to be published).
\bi{jos1}
A.~Waldron and G.~C.~Joshi, {\it Melbourne Preprint UM-P-92/60} (1992).\\
\bi{jos2}
G.~C.~Lassig and G.~C.~Joshi, {\it Melbourne Preprint UM-P-95/09} (1995).\\
A.~Ritz and G.~C.~Joshi, {\it Melbourne Preprint UM-P-95/69} (1995).
\bi{dav}
A.~J.~Davies and G.~C.~Joshi, \jxe{27}{3036}{86}.
\bi{odir}
S.~De Leo and K.~Abdel-Khalek, {\it Octonionic Dirac Equation}, 
Progr.~Theor.~Phys. (submitted).
\bi{itz}
C.~Itzykson and J.~B.~Zuber, {\it Quantum Field Theory} 
(McGraw-Hill, New York, 1985).
\bi{oqm}
S.~De Leo and K.~Abdel-Khalek, {\it Octonionic Quantum Mechanics and 
Complex Geometry}, Progr.~Theor.~Phys. (submitted).
\bi{math}
S.~Wolfram, {\it Mathematica} (Addison-Wesley Publishing Co., Redwood City, 
1991).

\eref

\end{document}